\documentclass{IEEEtran}
\usepackage{amssymb,subfigure,algorithm,bm,algpseudocode,dsfont,stfloats}
\usepackage{stfloats}
\usepackage[dvips]{graphicx}
\usepackage[cmex10]{amsmath}
\newcommand{\bi}{\begin{itemize}}
\newcommand{\ei}{\end{itemize}}
\newcommand{\beq}{\begin{equation}}
\newcommand{\eeq}{\end{equation}}
\newcommand{\bqn}{\begin{eqnarray*}}
\newcommand{\eqn}{\end{eqnarray*}}
\newcommand{\ba}{\begin{array}}
\newcommand{\ea}{\end{array}}
\newcommand{\bs}{\begin{small}}
\newcommand{\es}{\end{small}}
\newcommand{\nn}{\nonumber}

\providecommand{\norm}[1]{\lVert#1\rVert}

\newtheorem{theorem}{Theorem}
\newtheorem{lemma}{Lemma}
\newtheorem{remark}{Remark}

\begin{document}
\title{Asymptotic Properties of Likelihood Based Linear Modulation Classification Systems}
\author{Onur~Ozdemir\footnotemark{*},~\IEEEmembership{Member,~IEEE}, Pramod~K.~Varshney,~\IEEEmembership{Fellow,~IEEE}, Wei~Su,~\IEEEmembership{Fellow,~IEEE}, Andrew~L.~Drozd,~\IEEEmembership{Fellow,~IEEE}%
%\thanks{EDICS: NEED TO PUT EDICS!!!}
\thanks{O. Ozdemir and A. L. Drozd are with Andro Computational Solutions, 7902 Turin Road, Rome, NY 13440. P. K. Varshney is with Department of EECS, Syracuse University, Syracuse, NY 13244. W. Su is with U.S. Army CERDEC, Aberdeen Proving Ground, MD 21005. This work was supported by U.S. Army  contract W15P7T-11-C-H262.
Email:  \{oozdemir, adrozd\}@androcs.com, varshney@syr.edu, wei.su@us.army.mil}}
\markboth{IEEE TRANSACTIONS ON WIRELESS COMMUNICATIONS (DRAFT)}{}

\maketitle
\begin{abstract}
The problem of linear modulation classification using likelihood based methods is considered. Asymptotic properties of most commonly used classifiers in the literature are derived. These classifiers are based on hybrid likelihood ratio test (HLRT) and  average likelihood ratio test (ALRT) respectively. Both a single-sensor setting and a multi-sensor setting that uses a distributed decision fusion approach are analyzed. For a modulation classification system using a single sensor, it is shown that HLRT achieves asymptotically vanishing probability of error ($P_e$) whereas the same result cannot be proven for ALRT. In a multi-sensor setting using soft decision fusion, conditions are derived under which $P_e$ vanishes asymptotically. Furthermore, the asymptotic analysis of the fusion rule that assumes independent sensor decisions is carried out. 
\end{abstract}

%\newpage
\section{Introduction}
%In this paper, we analyze the probability of error ($P_e$) in a distributed modulation classification system.  We start with a single sensor and then extend the results to a general multi-sensor scenario. Furthermore, we derive performance bounds on $P_e$ using Chernoff and Bhattacharyya error exponents, and propose methods to compute these bounds using Monte Carlo sampling techniques. 
%We also derive the optimal fusion rule for a general multi-hypothesis decision fusion system and provide comparison results with the simpler majority fusion rule to gain further insight into the distributed modulation classification problem.

Automatic modulation classification (AMC) is a signal processing technique that is used to estimate the modulation scheme corresponding to a received noisy communication signal. It plays a crucial role in various civilian and military applications, e.g., this technique has been widely used in many communication applications such as spectrum monitoring and adaptive demodulation. 
%Recently, AMC has been adopted by cognitive radios to estimate the modulation scheme of an unknown noisy signal. The cognitive radios are usually implemented using the networked software-defined radio (SDR) platforms which demodulate signals using the estimated modulation schemes \cite{su_icnsc10_2}.
%The design of a modulation classifier essentially involves two steps: signal preprocessing and proper selection of the classification algorithm. Preprocessing tasks may include, but not limited to performing some or all of, noise reduction, estimation of carrier frequency, symbol period, and signal power, equalization, etc. Depending on the classification algorithm chosen in the second step, preprocessing tasks with different levels of accuracy are required; some classification methods require precise estimates, whereas others are less sensitive to the unknown parameters \cite{su_iet07}.
The AMC methods can be divided into two general classes (see the survey paper \cite{su_iet07}): 1) likelihood-based (LB) and 2) feature-based (FB) methods. In this paper, we focus on the former method which is based on the likelihood function of the received signal under each modulation scheme, where the decision is made using a Bayesian hypothesis testing framework. The solution obtained by the LB method is optimal in the Bayesian sense, i.e., it minimizes the probability of incorrect classification. In the last two decades, extensive research has been conducted on AMC methods, which are mainly limited to methods based on receptions at a single sensor (communication receiver). A detailed survey on the AMC techniques using a single sensor can be found in \cite{su_iet07}. For a single sensor tasked with AMC, the classification performance depends highly on the channel quality which directly affects the received signal strength. In non-cooperative communication environments, additional challenges  exist that further complicate the problem. These challenges stem from unknown parameters such as signal-to-noise ratio (SNR) and phase offset. In order to alleviate classification performance degradation in non-cooperative environments, network centric collaborative AMC approaches have been proposed in \cite{gian_milcom08, su_icnsc10, su_sj10, su_milcom11, su_globecom11}. It has been shown that the use of multiple sensors has the potential of boosting effective SNR, thereby improving the probability of correct classification. 

\begin{figure}[h]
\centering
\includegraphics[width=0.45\textwidth,height=!]{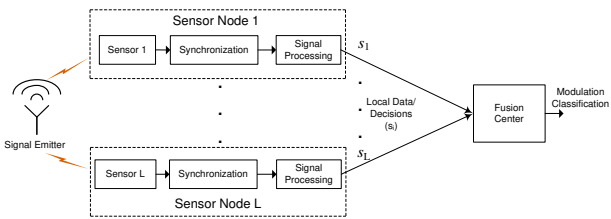}
\caption{Generic system model for a multi-sensor modulation classification system. $s_l$ is the decision/data of the $l$th sensor, where $l=1,\ldots,L$.}
\label{fig:sysmodel}
\end{figure}

In this paper, we focus on the likelihood based classification of linearly modulated signals, i.e., PSK and QAM signals. We notice that this problem is a composite hypothesis testing problem due to unknown signal parameters, i.e., uncertainty in the parameters of the probability density functions (pdfs) associated with different hypotheses. Various likelihood ratio based automatic modulation classification techniques have been proposed in the literature. An underlying assumption in all of these techniques is that each hypothesis has equally likely priors, in which case the classifiers reduce to maximum likelihood (ML) classifiers. These techniques take the form of a generalized likelihood ratio test (GLRT), an average likelihood ratio test (ALRT) or a  hybrid likelihood ratio test (HLRT). A thorough review of these techniques can be found in \cite{su_tsmc_11}. In the GLRT approach, all the unknown parameters are estimated using maximum likelihood (ML) methods and then a likelihood ratio test (LRT) is carried out by plugging in these estimates into the pdfs under both hypotheses. In addition to its complexity, GLRT has been shown to provide poor performance in classifying nested constellation schemes such as QAM  \cite{su_sarnoff05}. In the ALRT approach \cite{su_tsmc_11}, the unknown signal parameters are marginalized out assuming certain priors converting the problem into a simple hypothesis testing problem. In the HLRT approach \cite{su_tsmc_11}, the likelihood function (LF) is marginalized over the unknown constellation symbols and then the resulting average likelihood function (LF) is used to find the ML estimates of the remaining unknown parameters. These estimates are then plugged into the average LFs to carry out the LRT. Also, there are several variations of HLRT, which are called quasi HLRT (QHLRT), in which the ML estimates are replaced with other alternatives such as moment based estimators. We do not discuss the details here and refer the interested reader to \cite{su_tsmc_11} for further details. Our goal in this paper is to derive asymptotic (in the number of observations $N$) properties of modulation classification methods. We consider both single sensor and multiple sensor approaches. Although there has been extensive work on developing various methods for modulation classification, to the best of our knowledge, except for the work in \cite{mendel_tc_00}, there is no work in the literature that investigates asymptotic properties of modulation classification systems under single sensor or multi-sensor settings. In  \cite{mendel_tc_00}, the authors consider a coherent scenario where the only unknown variables are the constellation symbols. In this scenario, they analyze the asymptotic behavior of ML classifiers for linear modualtion schemes. Using Kolmogorov-Smirnov (K-S) distance, they show that the ML classification error probability vanishes as $N\rightarrow \infty$. Our contributions in this paper are as follows. We start with a single sensor system and analyze the asymptotic properties of two AMC scenarios: 1) coherent scenario with known signal-to-noise ratio (SNR), 2) non-coherent scenario with unknown SNR. Although the first scenario is the same as the one considered in \cite{mendel_tc_00}, we provide a much simpler proof which is then utilized to obtain the results for our second scenario. We analyze both HLRT and ALRT approaches. We do not consider GLRT due to its poor performance in classifying nested constellations. After analyzing single sensor approaches, we consider a multi-sensor setting as shown in Fig. \ref{fig:sysmodel}. Under this framework, we analyze a specific multi-sensor approach, namely distributed decision fusion for multi-hypothesis modulation classification where each sensor uses the LB approach to make its local decision. In this setting, there are $L$ sensors observing the same unknown signal. Each sensor employs its own LB classifier and sends it soft decision to a fusion center where a global decision is made. We analyze the asymptotic properties of ALRT and HLRT in this multi-sensor setting in the asymptotic region as $N\rightarrow \infty$ and $L\rightarrow \infty$. We also provide implications of large number of observations for the fusion rule at the fusion center.

The rest of the paper is organized as follows. In Section \ref{sec:sys}, we introduce the system model and lay out our assumptions. In Section \ref{sec:lb}, we formulate the likelihood-based modulation classification problem and summarize HLRT and ALRT approaches. We consider the single sensor case in Section \ref{sec:sing} and analyze the asymptotic probability of classification error under various settings. Similarly, the asymptotic probability of classification error in the multi-sensor case is analyzed in Section \ref{sec:mult}. We provide numerical results that corroborate our analyses in Section \ref{sec:res}. Finally, concluding remarks along with avenues for future work are provided in Section \ref{sec:conc}.

\section{System Model Assumptions}\label{sec:sys}

We consider a general linear modulation reception scenario with multiple receiving sensors assuming that the wireless communication channel between the unknown transmitter and each sensor undergoes flat block fading, i.e., the channel impulse response is $h(t)=a e^{j\theta} \delta(t)$ over the observation interval. 
%The received signal at a given sensor is expressed as
%\beq
%r(t) = a e^{j\theta'}s(t-\tau)e^{j2\pi f_c t} + w(t), \label{eq:1}
%\eeq 
%where $s(t)$ denotes the time-varying message signal; $\tau$ denotes the signal time offset; $f_c$ is the carrier frequency; $a$ and $\theta'$ are the channel gain and the channel phase, respectively; and $w(t)$ is the additive zero-mean circularly symmetric complex Gaussian noise with real and imaginary parts of variance $N_0/2$. 
After preprocessing, the received complex baseband signal at each sensor can be expressed as \cite{su_iet07}:
\beq
r(t) = s(t|\tilde{\mathbf{u}}) + v(t),\quad 0\leq t\leq NT \label{eq:rec_sig}
\eeq
\beq
s(t|\tilde{\mathbf{u}}) = a e^{j\theta} e^{j2\pi \Delta f t} \sum_{n=0}^{N-1} I_{n} g_{tx}(t-n T-\varepsilon T ), \label{eq:gen_sig}
\eeq
where $s(t)$ denotes the time-varying message signal; $\tilde{\mathbf{u}}$ represents the unknown signal parameter vector; $a$ and $\theta$ are the channel gain (or the signal amplitude) and the channel (or the signal) phase, respectively; $v(t)$ is the additive zero-mean white Gaussian noise; $g_{tx}(t)$ is the transmitted pulse; $T$ is the symbol period; $\{I_n\}$ is the complex information sequence, i.e., the constellation symbol sequence; and $\varepsilon$ and  $\Delta f$ represent residual time and frequency offsets, respectively. The constant $\varepsilon T$ represents the propagation time delay within a symbol period where $\varepsilon \in [0,1)$. Throughout the paper, we assume that $\varepsilon$ and  $\Delta f$ are perfectly known. Therefore, without loss of generality, we set $\varepsilon=\Delta f=0$. The representation in (\ref{eq:gen_sig}) has the implicit assumption that phase jitter is negligible. Without loss of generality, we further assume that the constellation symbols have unit power, i.e., $E[|I_n|^2]=1$, where $E[\cdot]$ denotes statistical expectation. Note that the unknown phase term denoted by $\theta$ in (\ref{eq:gen_sig}) subsumes both the unknown channel phase and unknown carrier phase. Similarly, the unknown signal amplitude $a$ subsumes the unknown signal amplitude as well as the unknown channel gain. 

After filtering the received signal with a pulse-matched filter $g_{rx}(t)$, and sampling at a rate of $Q/T$, where $Q$ is an integer, the following discrete-time obervation sequence is obtained \cite{gian_tc_98}: 
\beq
r_k = s_k(\tilde{\mathbf{u}}) + w_k \label{eq:rec_sig1}
\eeq
\beq
s_k(\tilde{\mathbf{u}}) = a e^{j\theta} \sum_{n=0}^{N-1} I_{n} g(kT/Q-n T), \label{eq:gen_sig1}
\eeq
where $g(t) = g_{tx}(t)*g_{rx}(t)$ with $*$ denoting the convolution operator, $r_k = r(t)*g_{rx}(t)|_{t=kT/Q}$, $w_k = v(t)*g_{rx}(t)|_{t=kT/Q}$, $N$ is the total number of observed information symbol, and $k = 0,\ldots,K-1$.  Note that $N =  K/Q$, i.e., there are $Q$ samples per symbol. For simplicity, we assume that $g_{tx}(t)$ is a rectangular pulse where $g(t) = 1,\;0\leq t \leq T$. We further assume $Q=1$ and $w_n$ is independent identically distributed (i.i.d.) circularly symmetric complex Gaussian noise with real and imaginary parts of variance $N_0/2$, i.e., $w_n\thicksim\mathcal{CN}(0,N_0)$. Our analysis in this paper can be easily generalized to other pulse shapes and cases where $Q>1$. Under these assumptions, the received observation sequence can be written as:
\beq
r_n = a e^{j\theta} I_n + w_n ,\quad n=0,\ldots,N-1. \label{eq:simp_rec}
\eeq
The above signal model is a commonly used model in modulation classification literature \cite{su_iet07,dobre_ciss04,dobre_twc09,silva_tc11}. Note that $a$, $\theta$, and $\{I_n\}_{n=1}^N$ are the unknown signal parameters. In a general modulation classification scenario, in addition to the unknown signal parameters, the noise power $N_0$ may also be unknown. In this case, the unknown parameter vector can be written as $\tilde{\mathbf{u}}=\left[a,\,\theta,\,N_0,\,\{I_{n}\}_{n=0}^{N-1}\right]$.

\section{Likelihood-based Linear Modulation Classification} \label{sec:lb}
Our goal throughout this paper is to gain insights into the modulation classification problem using the assumptions commonly made in the modulation classification literature. Suppose there are $S$ candidate modulation formats under consideration. Let $\mathbf{r}$ denote the observation vector defined as $\mathbf{r} := [r_0,\ldots,r_{N-1}]$ and $I_n^{(i)}$ denote the constellation symbol at time $n$ corresponding to modulation $i\in\{1,\ldots,S\}$. The conditional pdf of $\mathbf{r}$ conditioned on the unknown modulation format $i$ and the unknown parameter vector $\mathbf{u}$, i.e., the likelihood function (LF), is given by
\beq
p_i(\mathbf{r}|\mathbf{u}) = \frac{1}{(\pi N_0)^N} \exp\left(-\frac{1}{N_0} \sum_{n=0}^{N-1}|r_n- a e^{j\theta} I_n^{(i)}|^2\right). \label{eq:lf1}
\eeq
If the transmitted signal is an M-PSK signal, the constellation symbol set is given as $\mathcal{S}_P^M = \{e^{j2\pi m/M}|m=0,\ldots,M-1\}$ and $I_n^{(i)}\in\mathcal{S}_P^M$. Otherwise, if the transmitted signal is an M-QAM signal, the constellation symbol set is $\mathcal{S}_Q^M = \{b_me^{j\theta_m}|m=0,\ldots,M-1\}$ and $I_n^{(i)}\in\mathcal{S}_Q^M$\footnote{In certain cases, these sets can be rotated by some fixed phase, e.g., QPSK is represented as a rotated version of $\mathcal{S}_P^4$ by $e^{j\pi/4}$. This does not affect our results.}.

Note that the LF in (\ref{eq:lf1}) is parameterized by the modulation scheme under consideration and the only difference between conditional pdfs of different modulation schemes comes from the constellation symbols $I_n$. In a Bayesian setting, the optimal classifier in terms of minimum probability of classification error is the maximum a posteriori (MAP) classifier. If there is no \textit{a priori} information on probability of modulation scheme employed by the transmitter available, which is usually the case in a noncooperative environment, one can use a non-informative prior, i.e., each modulation scheme is assigned an identical prior probability. This is the assumed scenario in this paper. In this case, the optimal classifier takes the form of the maximum likelihood (ML) classifier. 

Let us first consider the HLRT approach, where the LF is averaged over the unknown constellation symbols $I_n$ and then maximized over the remaining unknown parameters. The modulation scheme that maximizes the resulting LF is selected as the final decision, i.e.,
\beq
\hat{i} = \arg\max_{i=1,\ldots,S} \left(
\max_{a,\theta,N_0} E_{I_n^{(i)}} \{p_i(\mathbf{r}|\mathbf{u})\},
\right) \label{eq:ml1}
\eeq
where $E_x[\cdot]$ denotes the expectation operator with respect to the random variable $x$, and $I_n^{(i)}$ is the unknown constellation symbol for modulation format $i$.  

In the ALRT approach, the unknown parameters are all marginalized out resulting in the marginal likelihood function which is used to make the final decision as
\beq
\hat{i} = \arg\max_{i=1,\ldots,S} E_{\mathbf{u}}\left\{p_i(\mathbf{r}|\mathbf{u})\right\}.  \label{eq:ml2}
\eeq
In the next section, we analyze the probability of classification error starting with a single sensor setting followed by a multi-sensor setting.
\section{Asymptotic Probability of Error Analysis: Single Sensor Case}\label{sec:sing}
\subsection{Scenario 1: Coherent Reception with Known SNR}
In this scenario, the only unknown variables are the data symbols $I_n$, $n=1,\dots,N$. In this case, without loss of generality, the received complex signal can be expressed as
\beq
r_n  = I_n + w_n, \quad n=1,\ldots,N,
\eeq
Assuming independent information symbols and white sensor noise, the LF averaged over the unknown constellation symbols under modulation format $i$ is given as
\beq
p_i(\mathbf{r}) := p(\mathbf{r}|H_i) = \prod_{n=1}^N p(r_n|H_i), \label{eq:model1}
\eeq
where
\begin{align}
p(r_n|H_i) &= E_{I_n^{(i)}} \{p(r_n|H_i,I_n^{(i)})\} \nn\\
&= \sum_{m=1}^{M_i}p(r_n|I_n^{m,(i)},H_i)p(I_n^{m,(i)}|H_i).
\label{eq:1}
\end{align}
In ($\ref{eq:1}$), $M_i$ and $I_n^{m,(i)}$ are the number of constellation symbols and the $m^{th}$ constellation symbol for modulation class $i$, respectively. In general, the constellation symbols are assumed to have equal \emph{a priori} probabilities, i.e., $p(I_n^{m,(i)}|H_i)=1/M_i$, which results in
\beq
p(r_n|H_i) = \frac{1}{M_i}\sum_{m=1}^{M_i}p(r_n|I_n^{m,(i)},H_i).\label{eq:model2}
\eeq
where
\beq
p(r_n|I_n^{m,(i)},H_i) = \frac{1}{\pi N_0}\exp\left(-\frac{1}{N_0} |r_n- I_n^{m,(i)}|^2\right)
\eeq
In this case, $p(r_n|H_i)$ in (\ref{eq:model2}) represents a complex Gaussian mixture model (GMM), or a complex Gaussian mixture distribution, with $M_i$ homoscedastic components where each component has identical occurrence probability (weight) $1/M_i$ as well as identical variance $N_0$, and the mean of each component is one of the unique constellation symbols in modulation format $i$. Let us revisit the generic expression for a complex GMM denoted by $f(r)$:
\beq
f(r) = \sum_{i=1}^M w_i \phi(r;\mu_i,\sigma_i^2) \label{eq:gmm1}
\eeq
where
\beq
\phi(r;\mu_i,\sigma_1^2) = \frac{1}{\pi \sigma_i^2}\exp\left(-\frac{|r-\mu_1|^2}{\sigma_i^2}\right) \label{eq:gmm2}
\eeq
We know that a GMM given by (\ref{eq:gmm1}) and (\ref{eq:gmm2}) is completely parameterized by the set $\{w_i,\mu_i,\sigma_i^2\}_{i=1}^M$ \cite{reynolds_springer08}.
\begin{remark}\label{remark1}
For a given modulation format $i$, the Gaussian mixture model (GMM) in (\ref{eq:model2}) is completely parameterized by the means of the components in the mixture, i.e., by the constellation symbol set $\mathcal{S}^{(i)} = \{I^{1,(i)},\ldots,I^{M_i,(i)}\}$. In other words, if $\mathcal{S}^{(i)}\neq \mathcal{S}^{(j)}$ then $p(r_n|H_i)$ and $p(r_n|H_j)$ represent two different GMMs.
\end{remark}

Let us now define the test statistics
\beq
\Lambda_i := -\frac{1}{N}\log p_i (\mathbf{r})= -\frac{1}{N}\sum_{n=1}^N\log p(r_n|H_i). \label{eq:def_delta}
\eeq
Then, the  ML classifier is given as
\beq
\hat{i} = \arg\min_{i=1,\ldots,S} \Lambda_i. \label{eq:ml_class}
\eeq
The classifier performance can be quantified in terms of the average probability of error ($P_e$) given as
\beq
P_e = \frac{1}{S}\sum_{i=1}^S P_e^i, \label{eq:avgpe}
\eeq
where $P_e^i$ is the probability of error under hypothesis $H_i$, i.e., given that modulation $i$ is the true modulation,
\beq
P_e^i = 1-P(\Lambda_i<\Lambda_j|H_i), \quad \forall j\neq i. 
\eeq
Now, we can state the following theorem which shows that the probability of error of the ML classifier vanishes asymptotically as $N\rightarrow\infty$. Note that the same result was also obtained in \cite{mendel_tc_00} using Kolmogorov-Smirnov (K-S) distance. Here, we provide a simpler proof than the one in \cite{mendel_tc_00}.

\begin{theorem}\label{th1}
The ML classifier in (\ref{eq:ml_class}) asymptotically attains zero probability of error for classifying digital amplitude-phase modulations regardless of the received SNR, i.e.,
\beq
\lim_{N\rightarrow\infty}P_e = 0.
\eeq
\end{theorem}
\begin{IEEEproof}
Suppose $H_i$ is the true hypothesis. In order to study the asymptotic $(N\rightarrow \infty)$ behavior of $\Lambda_j(\mathbf{r})$ under $H_i$, we follow the same technique as in \cite{sayeed_ipsn03} and write the following using the law of large numbers:
\begin{align}
\lim_{N\rightarrow \infty} \Lambda_j(\mathbf{r}) &= -E_i\left[\log p_j(r)\right]\\
&=E_i[\log (p_i(r)/p_j(r))]-E_i[\log p_i(r)]\\
&=D(p_i||p_j)+h_i(r) \label{eq:asymp}
\end{align}
where $E_i[\cdot]$ is the expectation under $H_i$, $D(p_i||p_j)$ is the Kullback-Leibler (KL) distance between $p_i$ and $p_j$ defined as $D(p_i||p_j) := E_i[\log (p_i(r)/p_j(r))]$, and $h_i({r})$ is the differential entropy  defined as $h_i(r) := -E_i[\log p_i(r)]$ \cite{cover_inftheory}.
Note that $h_i(r)$ is not a function of any modulation $j\neq i$. Therefore, under $H_i$, the only difference between test statistics $\Lambda_i$ and $\Lambda_j$ is the KL distance $D(p_i||p_j)\geq 0$, which is equal to zero if and only if $p_j=p_i$. Now, let us revisit the ML classification rule given in (\ref{eq:ml_class}),
\beq
\hat{j} = \arg\min_{j=1,\ldots,S} \lim_{N\rightarrow \infty} \Lambda_j(\mathbf{r}).\label{eq:ml_asymp}
\eeq
Since the second term in  (\ref{eq:asymp}) is independent of the test statistics under consideration, i.e., $\Lambda_j$, the only difference between different test statistics results from the the first term in (\ref{eq:asymp}), which is the KL distance $D(p_i||p_j)$. If $D(p_i||p_j)>0$ for $j\neq i$ and $D(p_i||p_j)=0$ for $j=i$, the ML classifier in (\ref{eq:ml_asymp}) will always decide
\beq
i = \hat{j} = \arg\min_{j=1,\ldots,S} \lim_{N\rightarrow \infty} \Lambda_j(\mathbf{r}). \label{eq:ml_asymp2}
\eeq
Therefore, (\ref{eq:ml_asymp2}) implies that perfect classification is obtained for any given SNR in the limit as $N\rightarrow \infty$ if and only if $D(p_i||p_j) >0,\; \forall j,i,\;j\neq i$. For digital phase-amplitude modulations, we know from (\ref{eq:model2}) that $p_i(r)$ represents a GMM and each modulation format corresponds to a unique GMM (see Remark \ref{remark1}). Therefore, $D(p_i||p_j) >0,\, \forall j,i,\;j\neq i$, which is the only condition needed for asymptotically vanishing error probability of the ML classifier.
\end{IEEEproof}

\subsection{Noncoherent Reception with Unknown SNR}\label{sec:sing_noncoh}
In this scenario, the received complex signal is expressed as
\beq
r_n  = ae^{j\theta}I_n + w_n, \quad n=1,\ldots,N.
\eeq
In this case, in addition to the unknown constellation symbols, there are three more unknown parameters which are channel amplitude ($a$), channel phase ($\theta$), and noise power ($N_0$). 
%We assume block fading channels, i.e., the unknown channel parameters stay the same over the observation interval. 
We will denote these additional unknown parameters in vector form as $\mathbf{u}=[a,N_0,\theta]$, where $a\in[0,\infty)$, $N_0\in[0,\infty)$ and $\theta \in [0,2\pi)$. 

Let us first consider the HLRT approach, where the unknown data symbols are marginalized out and the remaining unknown parameters are estimated using an ML estimator. In HLRT, these ML estimates are plugged into the likelihood function to perform the ML classification task. In practice, the complex channel gain $ae^{j\theta}$ can be either random or deterministic depending on the application. In deep-space communications, the channel gain can be assumed to be a deterministic time-independent constant \cite{hamkins_06}, whereas in urban wireless communications, the channel gain is often assumed to be random due to multipath effects resulting in fading. In fading channels, the duration over which the channel gain remains constant depends on the coherence time of the channel. Nevertheless, in HLRT, the channel gain is always treated as a deterministic unknown regardless of the application and ML estimation is employed to estimate $a$ and $\theta$. The resulting likelihood function for modulation $i$ can be written as
\beq
p_i(\mathbf{r},\hat{\mathbf{u}}_i) :=p_i(\mathbf{r}|H_i,\hat{\mathbf{u}}_i) =  \prod_{n=1}^N p(r_n|H_i,\hat{\mathbf{u}}_i), \label{noncoh_hlr}
\eeq
where
\beq
p(r_n|H_i,\hat{\mathbf{u}}_i) = \frac{1}{M_i}\sum_{m=1}^{M_i}p(r_n|H_i,\hat{\mathbf{u}}_i,I_n^{m,(i)}), \label{eq:lf11}
\eeq
\beq
\hat{\mathbf{u}}_i= \arg\max_{\mathbf{u}} \prod_{n=1}^N p(r_n|H_i,\mathbf{u}). \label{eq:mlest1}
\eeq
In order to be explicit, we re-write (\ref{eq:lf11}) as
\beq
p(r_n|H_i,\hat{\mathbf{u}}_i)= \frac{1}{M_i}\sum_{m=1}^{M_i}\frac{1}{\pi \hat{N}_{0,i}}\exp\left(-\frac{ |r_n- \hat{a}_ie^{j\hat{\theta}_i}I_n^{m,(i)}|^2}{\hat{N}_{0,i}}\right).\label{eq:lf12}
\eeq
From (\ref{eq:lf12}), we can see that $p(r_n|H_i,\hat{\mathbf{u}}_i)$ represents a complex GMM with $M_i$ homoscedastic components where each component has identical occurrence probability $1/M_i$ as well as identical variance $\hat{N}_{0,i}$, and the mean of each component is one of the unique constellation symbols in modulation format $i$ mutiplied by $\hat{a}_ie^{j\hat{\theta}_i}$.
%\begin{remark}\label{remark1}
%For a given modulation format $i$, the Gaussian mixture model (GMM) in (\ref{eq:lf12}) is completely parameterized by the means of the components in the mixture, i.e., by the constellation symbol set $\mathcal{S}^{(i)} = \{I^{1,(i)},\ldots,I^{M_i,(i)}\}$. In other words, if $\mathcal{S}^{(i)}\neq \mathcal{S}^{(j)}$ then $p(r_n|H_i)$ and $p(r_n|H_j)$ represent two different GMMs.
%\end{remark}

We can define the new test statistics which now includes the estimates of the unknown parameters as
\beq
\Lambda_i(\mathbf{r},\hat{\mathbf{u}}_i) := -\frac{1}{N}\log p_i (\mathbf{r}|\hat{\mathbf{u}}_i)= -\frac{1}{N}\sum_{n=1}^N\log p(r_n|H_i,\hat{\mathbf{u}}_i). \label{eq:def_delta2}
\eeq
Then (\ref{eq:mlest1}) can be equivalently written as
\beq
\hat{\mathbf{u}}_i= \arg\min_{\mathbf{u}} \Lambda_i(\mathbf{r},\mathbf{u}), 
\eeq
and the ML classifier is given as
\beq
\hat{i} = \arg\min_{i=1,\ldots,S} \Lambda_i(\mathbf{r},\hat{\mathbf{u}}_i). \label{eq:ml_class2}
\eeq
We start the analysis by making the following observations. In practice, there is always some $\emph{a priori}$ knowledge on the bounds of the unknown parameters $a$ and $N_0$. In other words, the search space for the maximization of the likelihood function with respect to $a$ and $N_0$ can be confined to $[0,A^U]$ and $[0,N^U]$, respectively, for some known $A^U$ and $N^U$. Regarding the unknown phase $\theta$, the search space depends on the modulation class that is under consideration. For M-PSK modulations, it suffices to limit the search space of $\theta$ to $[0,2\pi/M)$, because the likelihood function is a periodic function of $\theta$ with a period of $2\pi/M$. This is due to averaging over the unknown constellation symbols and rotation of the constellation map with respect to $\theta$, i.e., rotation of the constellation map by $2\pi/M$ results in the same constellation map as far as the likelihood function averaged over the constellation symbols is considered. Similarly, for M-QAM modulations, it suffices to limit the search space of $\theta$ to $[0,\pi/2)$ because of the same reasons as M-PSK modulations discussed earlier. We now make the following assumption which will simplify mathematical analysis. We assume that the unknown parameters $[a,N_0,\theta]$ lie in the interior region of the cube $[0,A^U]\times[0,N^U]\times[0,2\pi/M]$ for M-PSK or $[0,A^U]\times[0,N^U]\times[0,\pi/2]$ for M-QAM, respectively. Note that these assumptions are almost always satisfied in practice. Let us denote this closed Euclidean space as $\mathbb{U}:[0,A^U]\times[0,N^U]\times[0,\theta^U]$, where $\theta^U = 2\pi/M$ for M-PSK and $\theta^U = \pi/2$ for M-QAM.

\begin{lemma}\label{lem1}
Let $\mathcal{S}$ denote the set of PSK and QAM modulation classes. Define $p_i(r|\mathbf{u}_i) := p(r|H_i,\mathbf{u}_i)$. Let $i,j\in\cal{S}$, $\mathbf{u}_i\in\mathbb{U}_i$, $\mathbf{u}_j\in\mathbb{U}_j$. If $i\neq j$, then
\beq
D(p_i(r|\mathbf{u}_i)||p_j(r|\mathbf{u}_j))>0.
\eeq
\end{lemma}
\begin{IEEEproof}
See Appendix \ref{app:lem1}.
\end{IEEEproof}

The following theorem states that the probability of error of the HLRT classifier vanishes asymptotically as $N\rightarrow\infty$.  
\begin{theorem}\label{th2}
The ML classifier in (\ref{eq:ml_class2}) asymptotically attains zero probability of error for classifying digital amplitude-phase modulations regardless of the received SNR.
\end{theorem}
\begin{IEEEproof}
Suppose $H_i$ is the true hypothesis and $\mathbf{u}_i^*$ denotes the true value of the unknown parameter. We start by noting that the maximum likelihood estimator (MLE) is consistent under some mild regularity conditions \cite{berger_statinf}, which are satisfied by the likelihood functions of digital amplitude-phase modulations. In other words, if $H_i$ is the true hypothesis and $\mathbf{u}_ i^*$ is the true value of the unknown parameter $\mathbf{u}$, then
\beq
\mathbf{u}_i^*=\arg\min_{\mathbf{u}} \lim_{N\rightarrow\infty}\Lambda_i(\mathbf{r},\mathbf{u}).
\eeq
Under $H_i$,  we write the following using the law of large numbers
\beq
\lim_{N\rightarrow \infty} \Lambda_j(\mathbf{r},\hat{\mathbf{u}}_j) = -E_i\left[\log p_j(r|\hat{\mathbf{u}}_j)\right], \label{eq:th2}
\eeq
where $E_i[\cdot]$ denotes expectation with respect to $p(r|H_i,\mathbf{u}_i^*)$. Then, ($\ref{eq:th2}$) can be written as
\begin{align}
\lim_{N\rightarrow \infty} \Lambda_j(\mathbf{r},\hat{\mathbf{u}}_j)=E_i&[\log(p_i(r|\mathbf{u}_i^*)/p_j(r|\hat{\mathbf{u}}_j))]- \label{eq:asymp2_1}\\
&E_i[\log p_i(r|\mathbf{u}_i^*)] \nn\\
&\hspace{-0.85cm}=D(p_i(r|\mathbf{u}_i^*)||p_j(r|\hat{\mathbf{u}}_j))+h_i(r|\mathbf{u}_i^*) 	\label{eq:asymp2}
\end{align}
where the second term is the differential entropy of the true distribution defined as $h_i(r|\mathbf{u}_i^*) :=-E_i[\log p(r|H_i,\mathbf{u}_i^*)]$. The proof follows from Lemma \ref{lem1} and the same reasoning as in Theorem \ref{th1}. 
\end{IEEEproof}
From (\ref{eq:asymp2}), we can make the following observation. Under $H_i$ and the true parameter $\mathbf{u}_i^*$, 
\begin{align}
\hat{\mathbf{u}}_j&=\arg\min_{\mathbf{u}} \lim_{N\rightarrow\infty}\Lambda_j(\mathbf{r},\mathbf{u})\\
&=\arg\min_{\mathbf{u}}D(p_i(r|\mathbf{u}_i^*)||p_j(r|\mathbf{u})).
\end{align}
As $N\rightarrow\infty$, the MLE $\hat{\mathbf{u}}_j$ minimizes the KL distance between the true and the assumed distributions. This was actually observed by Akaike \cite{akaike_isit73} in the area of maximum likelihood estimation under misspecified models (see also \cite{white_econ82}). We should also emphasize that the consistency of the ML estimator is necessary for $P_e$ to vanish as $N\rightarrow\infty$ as otherwise one cannot deduce (\ref{eq:asymp2}) from (\ref{eq:asymp2_1}). As one would expect, the result in Theorem \ref{th2} is useful in practice only when the channel gain remains constant over a large observation interval. Channels that exhibit such a behavior include deep space communication channels as well as slowly varying fading channels.

Next, we consider a variation of the HLRT approach where, in addition to unknown data symbols, a subset of remaining unknown parameters are marginalized out. Then the maximization is carried over the remaining subset. Let $\mathbf{u}^0$ denote the subset of the unknown parameters that are marginalized out and $f_{\mathbf{U}^0}(\mathbf{u}^0)$ denote the joint \emph{a priori} distribution of $\mathbf{u}^0$. Let $\mathbf{u}^1$ denote the vector of the remaining unknown parameters over which the maximization is carried out. Then, the ML classifier is given as
\beq
\hat{i} = \arg\max_{i=1,\ldots,S} p_i(\mathbf{r}|\hat{\mathbf{u}}_i^1)\label{eq:hlrt_sub1}
\eeq
\beq
\hat{\mathbf{u}}_i^1 = \arg\max_{\mathbf{u}^1} p_i(\mathbf{r}|\mathbf{u}^1)
\eeq
where
\beq
p_i(\mathbf{r}|\mathbf{u}^1)=\int_{\mathbf{U}^0}p_i(\mathbf{r}|\mathbf{u}^1,\mathbf{u}^0)f_{\mathbf{U}^0}(\mathbf{u}^0)d\mathbf{u}^0.\label{eq:avdep}
\eeq
Since the unknowns $[a,N_0,\theta]$ stay constant over the observation interval, it is clear from (\ref{eq:avdep}) that the observations $r_n$ become dependent after averaging (conditional independence is no longer valid), i.e.,
\beq
p_i(\mathbf{r}|\mathbf{u}^1)\neq\prod_{n=1}^Np_i(r_n|\mathbf{u}^1).\label{eq:hlrt_sub4}
\eeq
Due to this dependence, the law of large number cannot be invoked. Therefore, these classifiers do not have provably vanishing $P_e$ in the asymptotic regime as $N\rightarrow\infty$. This is also the case for the ALRT approach where all the unknowns are marginalized out before classification. In practice, ALRT may be preferred over HLRT since the latter requires multi-dimensional maximization of the LF which is generally a non-convex optimization problem. In order to alleviate this problem, a suboptimal HLRT called quasi-HLRT (or QHLRT) was proposed in \cite{su_sarnoff05,dobre_twc09}, where the MLEs of the unknown parameters were replaced with moment based estimators. In general, QHLRT does not guarantee provably asymptotically vanishing $P_e$, since these estimators are generally not consistent.

%Next, we consider the multiple sensor scenario. 

\section{Asymptotic Probability of Error Analysis: Multi-Sensor Case}\label{sec:mult}
In this section, we consider a multi-sensor setting where each sensor transmits its soft decision to a fusion center where a global decision is made. We start our analyses assuming soft decision fusion where each sensor sends its unquantized local likelihood value to the fusion center. 

In a multiple sensor scenario, the set of unknown parameters $\{a,\theta,N_0\}$ corresponding to each sensor is independent from that of other sensors. However, care must be taken to analyze this scenario as the independence of these unknowns does not guarantee the independence of different sensor observations. In the following, we will investigate the multiple sensor scenario and derive conditions under which the asymptotic error probability goes to zero.

\subsection{Scenario 1: Coherent Reception with Known SNR}
We first consider the general case for the coherent and synchronous environment where there are $L$ sensors and each sensor $l$ $(l=1,\ldots,L)$ makes $N$ observations. Let us define the vector of observations for each sensor as $\mathbf{r}_l := [r_{n_{l_1}},\ldots, r_{n_{l_N}}]$, $l = 1,\ldots, L.$. We also define the set of indices for the complex information sequence that each sensor observes as
\beq
\mathcal{I}_l := \{n_{l_1},\ldots, n_{l_N}\},\quad l = 1,\ldots, L.
\eeq
Similar to (\ref{eq:model1})-(\ref{eq:model2}), the likelihood function at sensor $l$ is
\beq
p_i(\mathbf{r}_l) := p(\mathbf{r}_l|H_i) = \prod_{n\in\mathcal{I}_l} p(r_n|H_i), \label{eq:mult_coh}
\eeq
\beq
p(r_n|H_i) = \frac{1}{M_i}\sum_{m=1}^{M_i}p(r_n|I_n^{m,(i)},H_i).
\eeq
Let $p_i(\mathbf{r}_s)$ and $p_i(\mathbf{r}_t)$ denote two arbitrary likelihood functions for sensor $s$ and $t$, where $s \neq t$. Assuming independent sensor noises, it is important to see that $\mathbf{r}_s\thicksim p_i(\mathbf{r}_s)$ and $\mathbf{r}_t\thicksim p_i(\mathbf{r}_t)$ are independent if and only if
\beq
\mathcal{I}_s \cap \mathcal{I}_t = \emptyset. \label{eq:ind_cond1}
\eeq
The condition in (\ref{eq:ind_cond1}) is required for independence since data symbols are marginalized out in the likelihood function. We should note that the implicit assumption in (\ref{eq:ind_cond1}) is that the data symbols are i.i.d. in time which is a common assumption in communications literature. From (\ref{eq:ind_cond1}), we can deduce the general condition for independence. All sensor observations are independent (across sensors) if and only if
\beq
\bigcap_{l=1,\ldots,L} \mathcal{I}_l = \emptyset. \label{eq:ind_con}
\eeq
Physically, the condition in (\ref{eq:ind_con}) implies that sensor observations, or the underlying baseband symbol sequences, should not overlap in time to satisfy independence. This condition may or may not be realized in practice. One possible way of obtaining independent sensor observations is to send a pilot signal to each sensor initiating data collection and leave enough time between two consecutive pilot signals so that each sensor observes a different non-overlapping time window of the same signal. 

Suppose the condition in (\ref{eq:ind_con}) is satisfied. Let $p_i^0$ denote the likelihood function at the fusion center for modulation $i$ defined as
\beq
p_i^0 := p(\mathbf{r}_1,\ldots,\mathbf{r}_L|H_i) = \prod_{l=1}^L\prod_{n\in\mathcal{I}_l} p(r_n|H_i). 
\eeq
We can now define 
\begin{align}
\Lambda_i^0 :=-\frac{1}{LN}\log p_i ^0 &= -\frac{1}{N}\sum_{l=1}^L\log p(\mathbf{r}_l|H_i) \label{eq:sec_in}\\
&= -\frac{1}{LN}\sum_{l=1}^L\sum_{n\in\mathcal{I}_l} \log p(r_n|H_i). \nn
\end{align}
Note that the independence condition is necessary in order for the second equality in (\ref{eq:sec_in}) to hold. Then, the  ML classifier is given as
\beq
\hat{i} = \arg\min_{i=1,\ldots,S} \Lambda_i^0. \label{eq:ml_class3}
\eeq

\begin{theorem}\label{th3}
As $\sum_{l = 1}^L|\mathcal{I}_l|\rightarrow \infty$\footnote{$|\cdot|$ is the cardinality operator.},
the ML classifier in (\ref{eq:ml_class3}) achieves zero probability of error for classifying digital amplitude-phase modulations regardless of the received SNRs at sensors.
\end{theorem}
\begin{IEEEproof}
The proof follows the same steps as in Theorem \ref{th1} and is omitted here for brevity.
\end{IEEEproof}

%\begin{figure*}[!b]
%\underline{\hspace{\textwidth}}
%\beq
%p^A(\mathbf{r}_l|H_i) = C E_{\{I_n^{(i)}\}_{n=1}^N}\left\{
%\frac{1}{1+\frac{\Gamma}{N_0}\sum_{n=0}^{N-1} |I_n^{(i)}|^2}\exp\left(\frac{\frac{\Gamma}{N_0^2}|\sum_{n=0}^{N-1} (r_n  I_n^{(i)*})|^2}{1+\frac{\Gamma}{N_0}\sum_{n=0}^{N-1} |I_n^{(i)}|^2}- \frac{1}{N_0} \sum_{n=0}^{N-1} |r_n|^2\right)
%\right\} \label{eq:avlf1}
%\eeq
%\end{figure*}

\subsection{Noncoherent Reception with Unknown SNR}
In this scenario, the received complex signal at sensor $l$ can be expressed as
\beq
r_{n_l}  = a_le^{j\theta_l}I_{n_l} + w_{n_l}, \quad n_l \in \mathcal{I}_l.
\eeq
The vector of unknowns for sensor $l$ is $\mathbf{u}_l = [a_l,\theta_l,N_{0_l}]$. Let us first consider the HLRT approach where sensor $l$ computes its likelihood by first marginalizing over the unknown symbols $I_{n_l},\;n_l \in \mathcal{I}_l$, and then plugging in the MLE of $\mathbf{u}_l$. Let us define the vector of observations at the fusion center as $\mathbf{r_0}:=[\mathbf{r_1},\ldots,\mathbf{r}_L]$. Suppose that the independence condition in  (\ref{eq:ind_con}) is satisfied. Let $p_i^H(\mathbf{r}_0)$ denote the likelihood function at the fusion center for the HLRT given as 
\beq
p_i^H(\mathbf{r}_0) := p(\mathbf{r}_0|H_i,\hat{\mathbf{u}}_1,\ldots,\hat{\mathbf{u}}_L)=\prod_{l=1}^L\prod_{n_l\in\mathcal{I}_l} p(r_{n_l}|H_i,\hat{\mathbf{u}}_l). 
\eeq
Following the same reasoning as in the single sensor scenario, we can claim that $P_e\rightarrow 0$ as $N \rightarrow \infty$ using Theorem \ref{th1}.  However, the same result cannot be claimed for finite $N$ even when $L \rightarrow \infty$ due to different unknown parameters at different sensors. 

If a subset of unknowns are marginalized out in the HLRT approach (see Section \ref{sec:sing_noncoh} eqs. (\ref{eq:hlrt_sub1})-(\ref{eq:hlrt_sub4})), the distribution at the fusion center takes the following form:
\beq
p(\mathbf{r}_0|H_i,\hat{\mathbf{u}}_{1,(i)}^1,\ldots,\hat{\mathbf{u}}_{L,(i)}^1)=\prod_{l=1}^Lp(\mathbf{r}_l|H_i,\hat{\mathbf{u}}_{l,(i)}^1), \label{eq:noncoh_mult2}
\eeq
where $\hat{\mathbf{u}}_{l,(i)}^1$ denotes the ML estimate of the remaining unknown parameters of sensor $l$ under $H_i$, i.e.,
\beq
\hat{\mathbf{u}}_{l,(i)}^1 = \arg\max_{\mathbf{u}^1} p_i(\mathbf{r}_l|\mathbf{u}^1)
\eeq
where
\beq
p_i(\mathbf{r}_l|\mathbf{u}^1)=\int_{\mathbf{U}^0}p_i(\mathbf{r}_l|\mathbf{u}^1,\mathbf{u}^0)f_{\mathbf{U}^0}(\mathbf{u}^0)d\mathbf{u}^0.\label{eq:avdep}
\eeq
Then, the ML classifier is given as
\beq
\hat{i} = \arg\max_{i=1,\ldots,S} p(\mathbf{r}_0|H_i,\hat{\mathbf{u}}_{1,(i)}^1,\ldots,\hat{\mathbf{u}}_{L,(i)}^1) \label{eq:hlrt_mult_sub1}
\eeq
Similar to (\ref{eq:hlrt_sub4}), since the unknowns $[a_l,N_{0_l},\theta_l]$, $l=1,\ldots,L$ stay constant over the observation interval, it is clear from (\ref{eq:avdep}) that the observations $r_{n_l}$ become dependent after averaging, i.e.,
\beq
p_i(\mathbf{r}_l|\mathbf{u}^1)\neq\prod_{n_l\in\mathcal{I}_l} p_i(r_{n_l}|\mathbf{u}^1).
\eeq
Therefore, these classifiers do not have provably vanishing $P_e$ in the asymptotic regime as $N\rightarrow\infty$ due to dependence or as $L \rightarrow \infty$ due to different unknown parameters at different sensors. 

%Following the same result as in Section \ref{sec:sing_noncoh}, it is clear that these classifiers do not have provably vanishing $P_e$ in the asymptotic regime as $N\rightarrow\infty$ due to dependence or as $L \rightarrow \infty$. 

Let us now consider the ALRT approach where all the unknowns are marginalized out. Denote the joint \emph{a priori} distribution of $\mathbf{u}_l$ as$f_{\mathbf{U}}(\mathbf{u})$. Let $p_i^A(\mathbf{r}_0)$ denote the likelihood function at the fusion center for ALRT defined as
\beq
p_i^A(\mathbf{r}_0):=\prod_{l=1}^L p_A(\mathbf{r}_l|H_i) \label{eq:alrt_mult}
\eeq
where
\beq
p_A(\mathbf{r}_l|H_i) = \int_{\mathbf{U}}p(\mathbf{r}_l|H_i,\mathbf{u})f_{\mathbf{U}}(\mathbf{u})d\mathbf{u}. \label{eq:avlf2}
\eeq
Now, define the following
\beq
\Lambda_i^A := -\frac{1}{L}\log p_i ^A(\mathbf{r}_l) = -\frac{1}{L}\sum_{l=1}^L\log p_A(\mathbf{r}_l|H_i).
\eeq
The ML classifier is given as 
\beq
\hat{i} = \arg\min_{i=1,\ldots,S} \Lambda_i^A. \label{eq:ml_class4}
\eeq
For ALRT, we consider a special case where $N_0$ is known\footnote{When there is no non-stationary interference in the environment, $N_0$ corresponds to stationary sensor background noise power which can be accurately estimated using offline techniques.}, $a$ is Rayleigh distributed with $E[a^2] = \Gamma$, and $\theta$ is uniformly distributed over $[-\pi,\pi]$, i.e., $\theta\thicksim\cal{U}[-\pi,\pi]$. From \cite{su_iet07}, we can write the conditional pdf at sensor $l$ as in (\ref{eq:avlf1}) shown at the top of the page, where $C$ is a normalizing constant which is identical for all modulation classes. Note that the expectation $E_{\mathbf{I}^{(i)}}$ in (\ref{eq:avlf1}) requires summation over $M_i^N$ combinations of constellation sequences which may be computationally prohibitive for large $N$. Alternatively, (\ref{eq:avlf1}) can be computed by changing the order of averaging operations, i.e., by first averaging over the unknown constellation symbols followed by averaging over the unknown channel phase and the channel amplitude. This alternative approach does not result in a closed form expression, therefore, it needs to be computed by using numerical techniques.
\begin{figure*}%[hbp]
%\underline{\hspace{\textwidth}}
\beq
p^A(\mathbf{r}_l|H_i) = C E_{\mathbf{I}^{(i)}}\left\{
\frac{1}{1+\frac{\Gamma}{N_0} \norm{\mathbf{I}^{(i)}}^2}\exp\left(\frac{\frac{\Gamma}{N_0^2} \norm{\mathbf{I}^{(i)H}\mathbf{r}_l}^2}{1+\frac{\Gamma}{N_0}\norm{\mathbf{I}^{(i)}}^2}- \frac{\norm{\mathbf{r}_l}^2}{N_0} \right)
\right\}\label{eq:avlf1}
\eeq
\underline{\hspace{\textwidth}}
\end{figure*}
\begin{lemma}\label{lem2}
Let $\mathcal{S}$ denote the set of PSK and QAM modulation classes. Define $p_i^A(\mathbf{r}_l):=p_A(\mathbf{r}_l|H_i)$ as given in (\ref{eq:avlf1}). For $i,j\in\cal{S}$, if $i\neq j$ and $N>1$, then $D(p_i^A(\mathbf{r}_l)||p_j^A(\mathbf{r}_l))>0$.
\end{lemma}
\begin{IEEEproof}
See Appendix \ref{app:lem2}.
\end{IEEEproof}
\begin{theorem}\label{th4}
Suppose $N_0$ is known, $a$ is Rayleigh distributed, and $\theta$ is uniformly distributed over $[-\pi,\pi]$. Then the ML classifier in  (\ref{eq:ml_class4}) achieves zero probability of error as $L\rightarrow\infty$.
\end{theorem}
\begin{IEEEproof}
The proof follows from Lemma \ref{lem2} and the same method as in Theorem \ref{th1}.
\end{IEEEproof}
Theorem \ref{th4} ensures that asymptotically vanishing $P_e$ is guaranteed in the number of sensors if ALRT is used at each sensor provided that each sensor has independent observations, i.e., each sensor observes a non-overlapping time window of the transmitted signal. In other words, using a multi-sensor approach ensures asymptotically vanishing $P_e$ for ALRT which is not provably the case for a single sensor as explained in Section \ref{sec:sing_noncoh}.

\subsection{Fusion Rule}
In this section, we analyze the implications of the independence condition in (\ref{eq:ind_con}) for decision fusion based modulation classification. For finite number of observations ($N<\infty$), it is clear that if (\ref{eq:ind_con}) is not satisfied, there are sensors observing the same baseband sequence resulting in dependent observations due to averaging over unknown constellation symbols. If (\ref{eq:ind_con}) is not satisfied, even though each sensor noise is independent, the joint conditional distribution at the fusion center cannot be written as a product of individual conditional distributions, i.e.,
\beq
p_i(\mathbf{r}_1,\ldots,\mathbf{r}_L)\neq \prod_{l=1}^L p_i(\mathbf{r}_l).
\eeq
However, in the asymptotic regime as $N\rightarrow\infty$, we have the following theorem.
\begin{theorem}\label{th5}
Suppose there are two groups of $L$ sensors denoted as $\mathcal{G}$ and $\mathcal{G}'$ observing the same signal with unknown modulation. Suppose the sensors in $\mathcal{G}$ have arbitrary overlaps in their observations and the sensors in $\mathcal{G}'$ have no overlaps. Let $\mathbf{r}_l$ and $\mathbf{r}'_l$, $l=1,\ldots,L$ denote the observations from the sensors in $\mathcal{G}$ and $\mathcal{G}'$, respectively. Let $p_i(\mathbf{r}_l)$ ($p_i(\mathbf{r}'_l)$) denote the likelihood function of sensor $l$ ($l'$) under $H_i$ which represents either a coherent scenario with known SNR as in (\ref{eq:mult_coh})  or a noncoherent scenario with unknown SNR in the forms of HLR or ALR as in (\ref{noncoh_hlr}) or (\ref{eq:avdep}) or (\ref{eq:avlf2}). Suppose both groups use the same fusion rule to classify the unknown modulation given as:
\begin{align}
&\mathcal{G}_1:\quad \hat{i} = \arg\max_i \prod_{l=1}^L p_i(\mathbf{r}_l) \label{eq:fus1},\\
&\mathcal{G}'_1:\quad \hat{i} = \arg\max_i \prod_{l=1}^L p_i(\mathbf{r}'_l) \label{eq:fus2}.
\end{align}
Let $P_e$ and $P'_e$ denote the probabilities of classification error for the fusion rules in (\ref{eq:fus1}) and (\ref{eq:fus2}), respectively. As $N\rightarrow\infty$, we have the following result:
\beq
\lim_{N\rightarrow\infty} (P_e-P'_e)=0
\eeq
%
%As $N\rightarrow\infty$, each sensor obervation becomes independent regardless of any overlap in the observed baseband sequences, i.e.,
%\beq
%\lim_{N\rightarrow\infty}p_i(\mathbf{r}_1,\ldots,\mathbf{r}_L) =  \prod_{l=1}^L p_i(\mathbf{r}_l).
%\eeq
\end{theorem}
\begin{IEEEproof}
Sensor observations in $\mathcal{G}$ are dependent. This dependence results solely from overlapping sensor observations regardless of the scenario under consideration and regardless of which classification algorithm is employed (HLR or ALR). Suppose $H_i$ is the hypothesis under consideration. Let $\mathcal{M}_i$ denote the set of constellation symbols for modulation $i$ with $|\mathcal{M}_i| = M_i$; and $I_n$, $n = 1,\ldots,N$ denote the received constellation symbol sequence by an arbitrary sensor. Suppose $s_m^{(i)} \in \mathcal{M}_i$ and let $\mathds{1}_{s_m^{(i)}}(I_n)$ denote the indicator function defined as $\mathds{1}_{s_m^{(i)}}(I_n)=1$ if $I_n = s_m^{(i)}$ or $\mathds{1}_{s_m^{(i)}}(I_n)=0$  otherwise. Now, define
\beq
\Omega(s_m^{(i)}) := \sum_{n=1}^N\mathds{1}_{s_m^{(i)}}(I_n)
\eeq
which represents the number of occurences of $s_m^{(i)}$ in the received symbol sequence $\{I_1,\ldots,I_N\}$. Now, take the limit 
\beq
\lim_{N\rightarrow\infty}\frac{1}{N}\Omega(s_m^{(i)}) \overset{(a)}{=}E_{s_m^{(i)}}[\mathds{1}_{s_m^{(i)}}(I_n)] \overset{(b)}{=} \frac{1}{M_i} \label{eq:count}
\eeq
where $(a)$ results from applying the law of large numbers and $(b)$ results from the fact that each symbol in the constellation set $\mathcal{M}_i$ is equally likely. We can rewrite (\ref{eq:count}) as
\beq
\lim_{N\rightarrow\infty}\Omega(s_m^{(i)}) = \frac{N}{M_i}, \label{eq:mult_count}
\eeq
which implies that as $N\rightarrow\infty$, each constellation symbol $s_m^{(i)} \in \mathcal{M}_i$ has identical number of occurences $\frac{N}{M_i}$. 
%In other words, the observed symbol sequence $\{I_1,\ldots,I_N\}$ contains identical count of each of the constellation symbol in $\mathcal{M}_i$ as $N\rightarrow\infty$. 
Therefore, in the asymptotic regime ($N\rightarrow\infty$), each sensor observes equal number of different constellation symbols whether those symbols overlap across sensors or not. 

Now, consider sensor $l$ and let $I_{l_n}$ denote the $n$-th symbol received by sensor $l$. Note that $p_i(\mathbf{r}_l) = \prod_{k=1}^N p_i(r_{l_k}) $ is permutation invariant with respect to $\mathbf{r}_l=[r_{l_1},\ldots,r_{l_N}]$ (or $\{I_{l_1},\ldots,I_{l_N}\}$), because each $I_{l_n}$ is i.i.d. and background noise is white. In other words, $p_i(\mathbf{r}_l)$ is invariant to the order of the received symbol sequence $\{I_{l_1},\ldots,I_{l_N}\}$. Let us define a virtual sensor indexed by $l'$ and suppose that it observes a symbol sequence  $\{I_{l'_1},\ldots,I_{l'_N}\}$ that does not overlap with those observed by other sensors, i.e., $\{I_{l_1},\ldots,I_{l_N}\}$ and $\{I_{l'_1},\ldots,I_{l'_N}\}$ represents i.i.d. symbol sequences. As we let $N\rightarrow\infty$, the number of occurences of each symbol in $\{I_{l_1},\ldots,I_{l_N}\}$ and $\{I_{l'_1},\ldots,I_{l'_N}\}$ become identical from (\ref{eq:mult_count}). This implies that $\{I_{l'_1},\ldots,I_{l'_N}\}$ becomes a re-ordered version of $\{I_{l_1},\ldots,I_{l_N}\}$. In this case (as $N\rightarrow \infty$), the elements of the observation vector $\mathbf{r}_l$ can be re-ordered to form a new observation vector $\mathbf{r}_{l'}$ such that it represents noisy observations of the virtual symbol sequence $\{I_{l'_1},\ldots,I_{l'_N}\}$. It follows that, since $p_i(\mathbf{r}_l)$ is permutation invariant with respect to $\mathbf{r}_l$, we have the following equality as $N\rightarrow \infty$:
%\beq
%\lim_{N\rightarrow\infty}p_i(\mathbf{r}_l|\mathbf{I}_l)= p_i(\mathbf{r}_{l'}|\mathbf{I}_{l'})
%\eeq
\beq
p_i(\mathbf{r}_l)= p_i(\mathbf{r}_{l'}).
\eeq
Similarly, we can follow the same argument as above and show that $p_i(\mathbf{r}_l)= p_i(\mathbf{r}_{l'})$, $l=1,\ldots,L$. This implies that as $N\rightarrow \infty$,
\beq
\prod_{l=1}^L p_i(\mathbf{r}_l)= \prod_{l=1}^L p_i(\mathbf{r}_{l'}).
\eeq
Finally, the above equality implies that as $N\rightarrow \infty$.
\beq
\arg\max_i \prod_{l=1}^L p_i(\mathbf{r}_l) = \arg\max_i \prod_{l=1}^L p_i(\mathbf{r}'_l),\label{eq:fus_res}
\eeq 
which concludes the proof. 
%%where $\mathbf{I}_l = \{I_{l_1},\ldots,I_{l_N}\}$ and $\mathbf{I}_{l'} = \{I_{l'_1},\ldots,I_{l'_N}\}$.
%\begin{align}
%&\lim_{N\rightarrow\infty}p_i(\mathbf{r}_1,\ldots,\mathbf{r}_L) \nn\\
%&= \sum_{\mathbf{\mathcal{I}}_1}\ldots \sum_{\mathbf{\mathcal{I}}_L} p_i(\mathbf{r}_1,\ldots,\mathbf{r}_L|\mathbf{I}_1,\ldots,\mathbf{I}_L)p_i(\mathbf{I}_1,\ldots,\mathbf{I}_L)\nn\\
%&=\sum_{\mathbf{\mathcal{I}}_1}\ldots \sum_{\mathbf{\mathcal{I}}_L} p_i(\mathbf{r}_1|\mathbf{I}_1)\ldots p_i(\mathbf{r}_L|\mathbf{I}_L) p_i(\mathbf{I}_1,\ldots,\mathbf{I}_L)
%\end{align}
%
%????????????????????????????
%Furthermore, we note that $p_i(\mathbf{r}_l)$ is permutation invariant with respect to $\{I_1,\ldots,I_N\}$ since each $I_n$ is i.i.d. The result follows from independent sensor noises, uniform counts of different constellation symbols and permutation invariant nature of $p_i(\mathbf{r}_l)$.
\end{IEEEproof}
%\begin{remark}\label{rem1}
%The practical implication of Theorem \ref{th5} is that, for large $N$, each sensor decision can be treated as an independent random variable at the fusion center, and the optimal fusion rule for ML classification becomes
%\beq
%\hat{i} = \arg\max_i \sum_{l=1}^L \log p_i(\mathbf{r}_l).
%\eeq
%Practical $N$ values for the above fusion rule to be optimal will be provided by numerical results in Section \ref{sec:res} for different modulation classification scenarios.
%\end{remark}
The above result shows that as $N\rightarrow \infty$, we can always re-arrange the order of original observations and create an equivalent system with independent observations resulting in a new system having the same classification performance as the original one provided that both systems use the same fusion rule. 
\begin{remark}\label{rem1}
We know that the optimal fusion rule for $\mathcal{G}'$ which minimizes $P'_e$ is given as $\hat{i} = \arg\max_i \prod_{l=1}^L p_i(\mathbf{r}'_l)$. The practical implication of Theorem \ref{th5} is that, for large $N$, regardless of any overlap in the sensor observations, the fusion rule $\hat{i} = \arg\max_i \prod_{l=1}^L p_i(\mathbf{r}_l)$ will achieve the performance which is the best that can be achieved by a multi-sensor system with independent sensor observations. Practical $N$ values for which this performance can be achieved will be provided by numerical results in Section \ref{sec:res} for different modulation classification scenarios. 
%\beq
%\hat{i} = \arg\max_i \sum_{l=1}^L \log p_i(\mathbf{r}_l).
%\eeq
\end{remark}

In practice, it may be impossible to characterize the dependence in sensor observations as sensors may have arbitrary and unknown overlaps in their observations. In this case, the optimal fusion rule simply cannot be derived and the fusion rule that assumes independence becomes a natural choice. Theorem \ref{th5} provides an asymptotic performance guarantee for such a scenario.  

%So far, we have only considered soft-decision fusion, i.e., analog likelihood functions are fused to make a final decision. Following the same methodology as in the previous sections, similar analyses can be carried out for hard-decision fusion scenarios, i.e., quantized likelihood fusion such as binary or multi-bit decision fusion. However, we can claim that the conclusions will stay the same with the exception that the rate of convergence of $P_e$ to zero will be smaller (for the cases where $P_e \rightarrow 0$) with hard-decision fusion than that for soft-decision fusion. In fact, the rate of convergence will be an increasing function of the number of bits used by each sensor. Since the analysis is similar and space is limited, we omit the details in this paper. 

\section{Numerical Results}\label{sec:res}
\begin{figure}[h]
\centering
\includegraphics[width=0.4\textwidth,height=!]{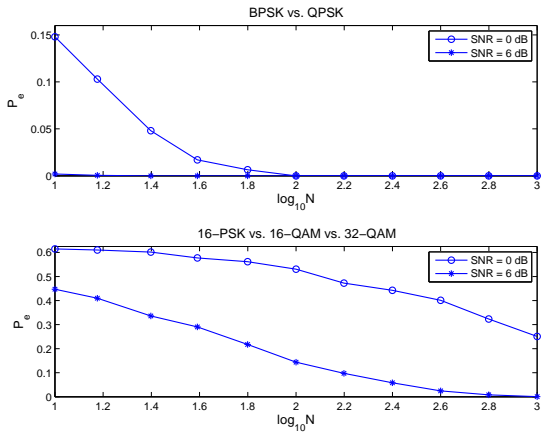}
\caption{Coherent scenario with known SNR. $P_e$ versus number of observations (N) under two different SNR regimes: $0$ dB and $6$ dB.}\label{fig:Pe_coh}
\end{figure}

\begin{figure}[h]
\centering
\includegraphics[width=0.4\textwidth,height=!]{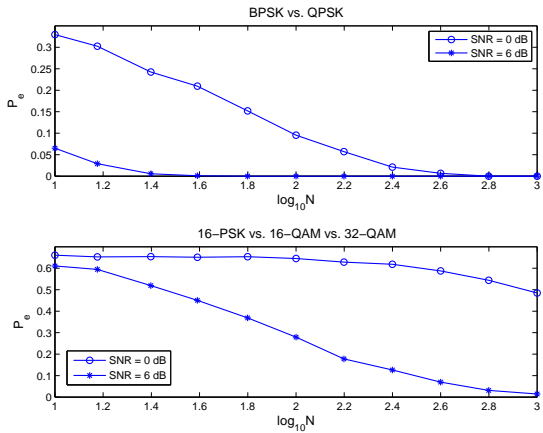}
\caption{Noncoherent scenario with unknown SNR. $P_e$ versus number of observations (N) under two different SNR regimes: $0$ dB and $6$ dB.}\label{fig:Pe_noncoh}
\end{figure}

In this section, we provide numerical results that corroborate our analyses in Sections \ref{sec:sing} and \ref{sec:mult}. First, we consider the single sensor case and investigate two classification scenarios: 1) binary classification of BPSK versus (vs.) QPSK, 2) 3-ary classification of 16-PSK vs.16-QAM vs. 32-QAM. Figures \ref{fig:Pe_coh} and  \ref{fig:Pe_noncoh} show $P_e$ versus number of observations $(N)$ under two different SNR regimes. The results are obtained using $2000$ Monte Carlo simulations. The difference between the two figures is that the former assumes a coherent scenario with known SNR whereas the latter assumes a noncoherent scenario with unknown SNR for which HLRT is used as the classifier. It is clear from both the figures that $P_e$ decreases monotonically as $N$ increases under both SNR regimes which support the analyses of Theorems \ref{th1} and \ref{th2}. As expected, the rate of decrease in $P_e$ is slower under $0$ dB SNR than that under $6$ dB SNR. Since Theorem \ref{th3} is an extension of Theorem \ref{th1} to a multi-sensor case, we do not provide additional results for that particular scenario.

\begin{figure}[h]
\centering
\includegraphics[width=0.35\textwidth,height=!]{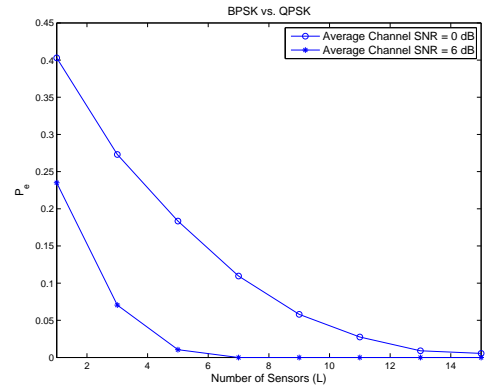}
\caption{ALRT with $N=4$ observations. $P_e$ versus number of sensors (L) under two different SNR regimes: $0$ dB and $6$ dB.}\label{fig:Pe_alrt}
\end{figure}

Fig. \ref{fig:Pe_alrt} demonstrates the performance of ALRT for classification of BPSK vs. QPSK with respect to number of sensors $(L)$ under two different SNR regimes. Each sensor receives a Rayleigh faded signal with an average channel SNR defined as $E[a^2]/N_0 = \Gamma/N_0$. The number of observations per sensor is set to $N=4$ . Similar to the previous cases, $2000$ Monte Carlo simulations are used to obtain the results. As stated by Theorem \ref{th4} and shown in Fig. \ref{fig:Pe_alrt}, $P_e$ decreases monotonically as $L$ gets larger regardles of the SNR regime. Furthermore, the rate of decrease in $P_e$ is slower for smaller SNR values as expected. 

\begin{figure}[h]
\centering
\includegraphics[width=0.35\textwidth,height=!]{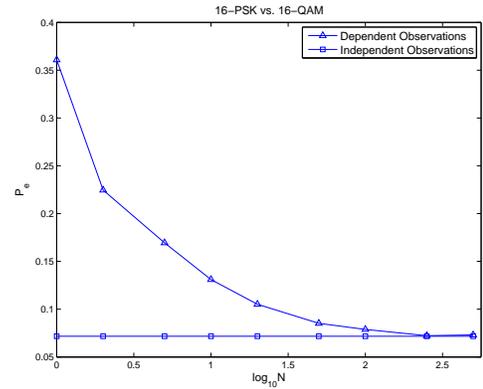}
\caption{$P_e$ with the fusion rule in (\ref{eq:fus1}) using dependent vs. independent observations (16-PSK vs. 16-QAM).}\label{fig:Pe_fus_binary}
\end{figure}

\begin{figure}[h!]
\centering
\includegraphics[width=0.35\textwidth,height=!]{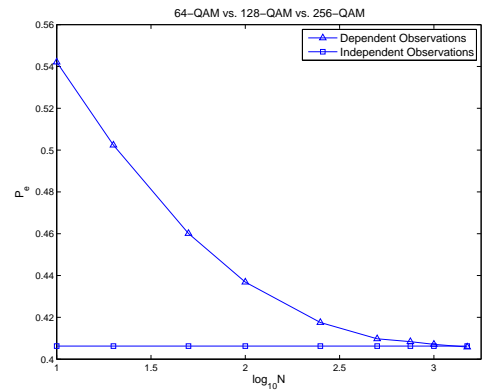}
\caption{$P_e$ for the fusion rule in (\ref{eq:fus1}) using dependent vs. independent observations (64-QAM vs. 128-QAM vs. 256-QAM).}\label{fig:Pe_fus_3ary}
\end{figure}

Finally, Figures \ref{fig:Pe_fus_binary} and \ref{fig:Pe_fus_3ary} illustrate how the fusion rule that assumes independent sensor decisions behaves asymptotically under $0$ dB SNR for two different classification scenarios: 1) binary classification of 16-PSK vs.16-QAM, 2) 3-ary classification of 64-QAM vs. 128-QAM vs. 256-QAM, respectively. Both figures assume coherent scenarios with known SNRs. In the figures, ``Independent Observations'' refers to the case where the condition in (\ref{eq:ind_con}) is satisfied, i.e., each sensor oberves a non-overlapping window of the signal, whereas ``Dependent Observations'' is the case where each sensor oberves the same window, i.e., there is complete overlap between sensor observations. Results are obtained using $10^4$ Monte Carlo simulations. In Fig. \ref{fig:Pe_fus_binary}, each marked point represents $L\times N = 1000$ observations and those points correspond to $N = \{1,2,5,10,20,50,100,250,500\}$ resulting in $L = \{1000,500,200,100,50,20,10,4,2\}$. When sensor observations are independent, $P_e$ is identical for all the points where $L\times N$ is constant. This is shown in both figures under ``Independent Observations'' case. It is clear from Fig. \ref{fig:Pe_fus_binary} that as $N$ grows, the performance of both systems converge supporting the analysis in Theorem \ref{th5}. For this particular scenario, when $N=250$ and $L=4$, the classification performance of the system with dependent observations is almost identical to that with independent observations where both fusion rules assume independent observations. In Fig. \ref{fig:Pe_fus_3ary}, each marked point represents $L\times N = 3000$ observations and those points correspond to $N = \{10,20,50,100,250,500,750,1000,1500\}$ resulting in $L = \{300,150,60,30,12,6,4,3,2\}$. For this scenario, when $N=1000$ and $L=3$, the classification performance of the system with dependent observations is almost identical to that with independent observations. We note that the convergence of the former scenario in Fig. \ref{fig:Pe_fus_binary} is faster than the latter in Fig. \ref{fig:Pe_fus_3ary}. This is due to the difference between cardinalities of constellation sets under consideration. Modulations with larger constellation sets require more number of observations for the mixing in (\ref{eq:count}) to take place. Therefore, practical $N$ values for which the two systems that use the same fusion rule behave identical is dependent on the classification scenario under consideration. 

\section{Conclusion}\label{sec:conc}
In this paper, we have investigated asymptotic behavior of LB modulation classification systems under two different scenarios: 1) coherent reception with known SNR, and 2) noncoherent reception with unknown SNR. Both a single-sensor setting and a multi-sensor setting that uses a distributed decision fusion approach are analyzed. In a single-sensor setting, it has been shown that $P_e$ vanishes asymptotically in the number of observations $(N)$ under coherent reception with known SNR. Under noncoherent reception with unknown SNR, HLRT achieves perfect classification, i.e., $P_e\rightarrow\infty$, in the asymptotic regime as $N\rightarrow\infty$, whereas this is not provably the case for ALRT. This property of HLRT is due to consistency of the ML estimator as well as statistical independence of data symbols in time. In a multi-sensor setting, under the assumption of independent sensor observations, it has been shown that perfect classification is achieved, i.e., $P_e\rightarrow\infty$, in the asymptotic regime as the number of sensors $L\rightarrow\infty$ provided that each sensor employs ALRT regardless of the number of observations ($N$). However, this is not provably the case when each sensor employs HLRT using a finite number of samples ($N<\infty$). Finally, the asymptotic analysis of the fusion rule that assumes independent sensor observations is carried out. It has been shown that this fusion rule asymptotically achieves the same performance as the best that can be achieved by a system employing independent sensor observations. The asymptotic results derived in this paper have practical implications in that they provide design guidelines as to which LB classification method should be selected for the specific scenario under consideration. Furthermore, they provide theoretical asymptotic performance guarantees for practical systems, which, otherwise, would be unknown.

As a future work, it would be interesting to investigate the case where each sensor makes hard decisions, i.e., quantized likelihoods are sent to the fusion center, instead of soft decisions (analog likelihoods) as assumed in this paper, and the fusion center employs hard decision fusion for modulation classification. We can conjecture that, under independent identical quantizer assumptions, one would obtain similar asymptotic results as for the soft decision fusion analyzed in this paper. Nevertheless, a rigorous treatment would be useful. Furthermore, we would like to incorporate additional unknown signal parameters such as frequency and time offsets into the signal model for similar asymptotic analyses in the future. 

\appendices
\section{Proof of Lemma \ref{lem1}}\label{app:lem1}
It is sufficient to show that if $i\neq j$, then $p(r|H_i,\mathbf{u}_i)$ and $p(r|H_j,\mathbf{u}_j)$ are not identical distributions for any $\mathbf{u}_i$, $\mathbf{u}_j$. We note from (\ref{eq:lf12}) that each $p(r|H_i,\mathbf{u}_i)$ is a complex GMM with $M_i$ components where each component has the same occurrence probability $1/M_i$, i.e., 
\begin{align}
p_i(r|\mathbf{u}_i) &=  \frac{1}{M_i}\sum_{m=1}^{M_i}p(r|H_i,\mathbf{u}_i,I^{m,(i)}) \nn\\
&= \frac{1}{M_i \pi}\sum_{m = 1}^{M_i }\frac{1}{N_{0,i}}\exp\left(\frac{|r-a_i e^{j\theta_i} I^{m,(i)}|^2}{N_{0,i}}\right) \label{eq:lemma1}
\end{align}
If the transmitted signal is an M-PSK signal, then $I^{m,(i)}\in\mathcal{S}_P^M$. Otherwise, if the transmitted signal is an M-QAM signal, then $I^{m,(i)}\in\mathcal{S}_Q^M$. From (\ref{eq:lemma1}), the mean value of each component in the GMM corresponds to a unique constellation symbol (in the constellation map of modulation format $i$) scaled by $a_i$ and rotated by $\theta_i$. The variance of each component is $N_{0,i}$. For different modulation classes $i$ and $j$, there are two cases to be considered: 
\begin{itemize}
\item[i)] Case-1: Modulations $i$ and $j$ represent two modulation classes with different number of constellation symbols. In this case, $p_i(r|\mathbf{u}_i)$ and $p_j(r|\mathbf{u}_j)$ represent two GMMs with different number of components, i.e., $M_i \neq M_j$. Therefore, $p_i(r|\mathbf{u}_i)$ and $p_j(r|\mathbf{u}_j)$ are not identical distributions and, hence, $D(p_i(r|\mathbf{u}_i)||p_j(r|\mathbf{u}_j))>0$.
\item[ii)] Case-2: Modulations $i$ and $j$ represent two modulation classes with the same number of constellation symbols. In this case, one of the modulation classes is M-PSK and the other is M-QAM. Suppose modulations $i$ and $j$  represent M-PSK and M-QAM, respectively. In this case, the mean value of each component in the GMM is given by $\mu_{i,(m)}\in\mathcal{S'}_P^M=\{a_i e^{j(2\pi m/M+\theta_i)}|m=0,\ldots,M-1\}$ and $\mu_{j,(m)}\in\mathcal{S'}_Q^M=\{a_j b_me^{j(\theta_m+\theta_j)}|m=0,\ldots,M-1\}$. We know from M-QAM constellation symbol set that there exist $m_1$ and $m_2$ such that $b_{m_1}\neq b_{m_2}$. In order for $p_i(r|\mathbf{u}_i)$ and $p_j(r|\mathbf{u}_j)$ to be identical, the following condition should be satisfied:
\beq
a_i e^{j(2\pi m/M+\theta_i)} = a_j b_me^{j(\theta_m+\theta_j)},\quad m=0,\ldots,M-1.\label{eq:cond11}
\eeq
\end{itemize}
Now suppose $p_i(r|\mathbf{u}_i)$ and $p_j(r|\mathbf{u}_j)$ are identical and consider $m_1$ and $m_2$ such that $b_{m_1}\neq b_{m_2}$. Then, from (\ref{eq:cond11}), we can write $a_i e^{j(2\pi m_1/M+\theta_i)} = a_j b_{m_1}e^{j(\theta_{m_1}+\theta_j)}$, which implies that $a_i/a_j = b_{m_1}$. Since $p_i(r|\mathbf{u}_i)$ and $p_j(r|\mathbf{u}_j)$ are identical, we can also write from (\ref{eq:cond11}) that $a_i e^{j(2\pi m_2/M+\theta_i)} = a_j b_{m_2}e^{j(\theta_{m_2}+\theta_j)}$ implying that $a_i/a_j = b_{m_2}$, which is a contradiction, because $b_{m_1}\neq b_{m_2}$. Then, $p_i(r|\mathbf{u}_i)$ and $p_j(r|\mathbf{u}_j)$ must be different GMMs, therefore, $D(p_i(r|\mathbf{u}_i)||p_j(r|\mathbf{u}_j))>0$. \QED

\section{Proof of Lemma \ref{lem2}}\label{app:lem2}
We drop the sensor index $l$ for simplicity of the presentation. There are three cases to be considered:
\begin{itemize}
\item[i)] Case-1: Modulations $i$ and $j$ represent two different PSK modulations, i.e., $I_n^{(i)}\in\mathcal{S}_P^{M_i}=\{e^{j2\pi m/M_i}|m=1,\ldots,M_i\}$, where $M_i = 2^{k_i}$, $k_i\in\mathbb{N}$. First, suppose $N=1$. Then, under $H_i$, (\ref{eq:avlf1}) becomes
%\vspace{-.5cm}
\begin{align}
&\hspace{-0.25cm}p^A(r|H_i)/C \nn\\
&= E_{I^{(i)}}\left\{
\frac{1}{1+\frac{\Gamma}{N_0}|I^{(i)}|^2}\exp\left(\frac{\frac{\Gamma}{N_0^2}|I^{(i)*}r|^2}{1+\frac{\Gamma}{N_0}|I^{(i)}|^2}- \frac{|r|^2}{N_0}\right)
\right\}\nn\\
&=\frac{1}{M_i}\sum_{m=1}^{M_i}  \frac{1}{1+\frac{\Gamma}{N_0}|I^{m,(i)}|^2}\exp\left(\frac{\frac{\Gamma}{N_0^2}|I^{m,(i)}|^2|r|^2}{1+\frac{\Gamma}{N_0}|I^{m,(i)}|^2}- \frac{|r|^2}{N_0}\right)\nn\\
&\stackrel{(a)}{=}\frac{1}{1+\frac{\Gamma}{N_0}}\exp\left(-\frac{|r|^2}{\Gamma+N_0}\right) \label{eq:N1}\
\end{align}
where $(a)$ follows from $E[|I^{(i)}|^2]=1$ and each symbol being equally likely, which implies that $|I^{m,(i)}|^2=1,\,\forall m$. We note that (\ref{eq:N1}) is independent of $H_i$. Therefore, $p^A(r|H_i)=p^A(r|H_j)$ which impies that $D(p_i^A(r)||p_j^A(r))=0$ for $N=1$. Now suppose $N>1$. In order to show that $D(p_i^A(\mathbf{r})||p_j^A(\mathbf{r}))>0$, it suffices to show that there exists an $\mathbf{r}_0$ such that  $p^A(\mathbf{r}_0|H_i)\neq p^A(\mathbf{r}_0|H_j)$. Let us set $\mathbf{r}_0 = \mathbf{1}$ (vector of ones) and write (\ref{eq:avlf1}) as
\begin{align}
&\hspace{-0.25cm}p^A(\mathbf{1}|H_i)/C \nn \\
&\hspace{-0.25cm}= E_{\mathbf{I}^{(i)}}\left\{
\frac{1}{1+\frac{\Gamma}{N_0}\norm{\mathbf{I}^{(i)}}^2}\exp\left(\frac{\frac{\Gamma}{N_0^2}\norm{\mathbf{I}^{(i)*}\mathbf{1}}^2}{1+\frac{\Gamma}{N_0}\norm{\mathbf{I}^{(i)}}^2}- \frac{\norm{\mathbf{1}}^2}{N_0}\right)
\right\}\nn\\
&\hspace{-0.25cm}=\frac{e^{-\frac{N}{N_0}}}{M_i^N}\sum_{m_1=1}^{M_i}\ldots\sum_{m_N=1}^{M_i} \frac{1}{1+\frac{\Gamma}{N_0}\sum\limits_{k=1}^{N}|I^{m_k,(i)}|^2}\nn\\
&\hspace{3.75cm}\exp\left(\frac{\frac{\Gamma}{N_0^2}|\sum\limits_{k=1}^{N}I^{m_k,(i)}|^2}{1+\frac{\Gamma}{N_0}\sum\limits_{k=1}^{N}|I^{m_k,(i)}|^2}\right)\nn\\
&\hspace{-0.25cm}=\frac{e^{-\frac{N}{N_0}}}{M_i^N(1+\frac{N\Gamma}{N_0})}\sum_{m_1=1}^{M_i}\ldots\sum_{m_N=1}^{M_i} \exp\left(\frac{\frac{\Gamma}{N_0^2}|\sum\limits_{k=1}^{N}I^{m_k,(i)}|^2}{1+\frac{N\Gamma}{N_0}}\right)\nn\\
&\hspace{-0.25cm}=\frac{e^{-\frac{N}{N_0}}}{M_i^N(1+\frac{N\Gamma}{N_0})}\sum_{m_1=1}^{M_i}\ldots\sum_{m_N=1}^{M_i} \nn\\
&\hspace{3.25cm} \exp\left(\frac{\frac{\Gamma}{N_0}}{N_0+N\Gamma}\sum\limits_{k=1}^{N}|I^{m_k,(i)}|^2\right)\nn\\
&\hspace{1.25cm}\exp\left(\frac{\frac{2\Gamma}{N_0}}{N_0+N\Gamma}\sum\limits_{k=1}^{N}\sum\limits_{h>k}^N\mathcal{R}\{I^{m_k,(i)*}I^{m_h,(i)}\}\right)\nn\\
&\hspace{-0.25cm}=\frac{e^{-\frac{N(N_0+N\Gamma-\Gamma)}{N_0(N_0+N\Gamma)}}}{M_i^N(1+\frac{N\Gamma}{N_0})}\sum_{m_1=1}^{M_i}\ldots\sum_{m_N=1}^{M_i} 
\nn\\
&\hspace{.75cm}\exp\left(\frac{\frac{2\Gamma}{N_0}}{N_0+N\Gamma}\sum\limits_{k=1}^{N}\sum\limits_{h>k}^N\mathcal{R}\{I^{m_k,(i)*}I^{m_h,(i)}\}\right) \nn\\
&\hspace{-0.25cm}=\frac{e^{-\frac{N(N_0+N\Gamma-\Gamma)}{N_0(N_0+N\Gamma)}}}{M_i^N(1+\frac{N\Gamma}{N_0})}\sum_{m_1=1}^{M_i}\ldots\sum_{m_N=1}^{M_i} 
\nn\\
&\hspace{0.25cm}\exp\left(\frac{\frac{2\Gamma}{N_0}}{N_0+N\Gamma}\sum\limits_{k=1}^{N}\sum\limits_{h>k}^N\cos\left(2\pi(m_h-m_k)/M_i\right)\right) \label{eq:N2}
\end{align}
where $\mathcal{R}\{\cdot\}$ denotes the real part of a complex number. We note that, for fixed $N>1$, (\ref{eq:N2}) cannot be reduced to a constant that is independent of $M_i$, i.e., $H_i$. In other words, for each $M_i = 2^{k_i}$, $k_i\in\mathbb{N}$, (\ref{eq:N2}) will result in a different value. Therefore, $p^A(\mathbf{1}|H_i)\neq p^A(\mathbf{1}|H_j)$ which implies that $D(p_i^A(\mathbf{r})||p_j^A(\mathbf{r}))>0$ for $N>1$.
\item[ii)] Case-2: Modulations $i$ and $j$ represent two QAM modulations, i.e., $I_n^{(i)}\in\mathcal{S}_Q^{M_i}=\{b_me^{j\theta_m}|m=1,\ldots,M_i\}$. Using the above methodology used in Case-1, we can show that $p_i^A(\mathbf{0})\neq p_j^A(\mathbf{0})$ for $N\geq1$ where $\mathbf{0}$ denotes vector of zeros. Details are omitted for the sake of brevity. Therefore, $D(p_i^A(\mathbf{r})||p_j^A(\mathbf{r}))>0$ for $N\geq1$.
\item[iii)] Case-3: Modulations $i$ and $j$ represent PSK and QAM modulations, respectively. In this case, similar to the above, we can show that $p_i^A(\mathbf{0})\neq p_j^A(\mathbf{0})$ for $N\geq1$. Details are omitted for the sake of brevity. Therefore, $D(p_i^A(\mathbf{r})||p_j^A(\mathbf{r}))>0$ for $N\geq1$. \QED
\end{itemize}

%If $N=1$, since $\theta$ is uniformly distributed, $r$ is uniformly distributed on the circle of radius $a$ regardless of which PSK symbol is transmitted. Therefore, $D(p_i^A(r)||p_j^A(r))=0$. If $N>1$, $\mathbf{r}$ is distributed non-identically on the circle of radius $a$ under modulations $i$ and $j$. This can be verified by setting $\mathbf{r} = \mathbf{1}$ (vector of ones) and plugging in (\ref{eq:avlf1}) for different PSK modulations which results in $p_i^A(\mathbf{1})\neq p_j^A(\mathbf{1})$. Now, suppose $i$ and $j$ represent two QAM modulations. It is clear from (\ref{eq:avlf1}) that $p_i^A(\mathbf{0})\neq p_j^A(\mathbf{0})$ for $N\geq1$ where $\mathbf{0}$ denotes vector of zeros. If $i$ and $j$ represent PSK and QAM modulations, respectively, we can again see from (\ref{eq:avlf1}) that $p_i^A(\mathbf{0})\neq p_j^A(\mathbf{0})$ for $N\geq1$.

\bibliographystyle{IEEEtran}
\bibliography{Journal,Conf,Book}

\end{document}